\documentclass[12pt]{article}
\pdfoutput=1
\usepackage{textcomp,xcolor}
\usepackage{amsmath,amssymb,latexsym,dsfont,soul}
\usepackage{slashed}
\usepackage[utf8]{inputenc}
\usepackage[german,english]{babel}
\usepackage{graphicx,floatflt,float,placeins,graphbox}
\usepackage{makeidx}
\usepackage[font=small,labelfont=bf]{caption}
\usepackage{nicefrac}
\usepackage{subfigure, multicol}
\usepackage{array, multirow, bigdelim}
\usepackage[Sonny]{fncychap}
\usepackage{fancyhdr}
\usepackage{lipsum}
\usepackage{etoolbox}
\usepackage[nomessages]{fp}
\usepackage{cite}
\usepackage{import}
\graphicspath{ }
\allowdisplaybreaks

  \def\cO{{\cal O}}

\newcommand{\TeV}{\text{TeV}}
\newcommand{\GeV}{\text{GeV}}

\newcommand{\vir}{~ ,}
\newcommand{\pun}{~ .}
\newcommand{\gsim}{\stackrel{>}{_\sim}}
\newcommand{\sst}{\scriptscriptstyle}
\newcommand{\sSM}{{\sst\rm SM}}
\newcommand{\MSSM}{{\rm \sst MSSM}}

\newcommand{\order}[1]{\cO \left(  #1 \right)}
\newcommand{\ba}[1]{\begin{array}{#1}}
\newcommand{\ea}{\end{array}}
\newcommand{\bea}{\begin{eqnarray}}
\newcommand{\eea}{\end{eqnarray}}
\newcommand{\be}{\begin{equation}}
\newcommand{\ee}{\end{equation}}
\newcommand{\eq}{Eq.~}

\newcommand{\Ref}{Ref.~}

\newcommand{\interline}{&&&&&\\[-11pt] \hline &&&&&\\[-10pt]}
\newcommand{\interlineB}{&&&&&\\[-11pt]}
\newcommand{\doubleinterline}{&&&&&\\[-11pt] \hline &&&&&\\[-13pt] \hline &&&&&\\[-15pt]}
\newcommand{\mS}{m_{S}}
\newcommand{\mG}{m_{G}}

\newcommand{\bS}{\beta}
\newcommand{\mHu}{m_{H_u}}
\newcommand{\mHd}{m_{H_d}}
\newcommand{\mHud}{m_{H_{u,d}}}

\newcommand{\mQU}{m_{Q_3,u_3}}

\newcommand{\At}{A_t}
\newcommand{\yt}{y_t}
\newcommand{\gt}{g_t}
\newcommand{\mt}{m_t}
\newcommand{\mh}{m_h}

\newcommand{\mtp}{v_F}

\newcommand{\gA}{g_{1,2,3}}

\newcommand{\rt}{r_t}
\newcommand{\rA}{\{ r_{t,Q,u,H_u,H_d} \}}

\newcommand{\rmu}{r_\mu}
\newcommand{\rb}{r_b}
\newcommand{\rmb}{r_{\mu,b}}
\newcommand{\ltr}{\lambda_{\rm thr}}

\begin{document}

\begin{flushright}
ZU-TH-32/17
\end{flushright}
\vspace{0.3cm}

\begin{center}
{\LARGE\bf
On the tuning in the $(m_h, m_t)$ plane: 
Standard Model criticality vs. High-scale SUSY \\
}

\bigskip\vspace{0.8cm}{
{\large \mbox{Gino Isidori$^{a}$ and Andrea Pattori$^{a,b}$} }
} \\[8mm]
  {\em $(a)$  Physik-Institut, Universit\"at Z\"urich, CH-8057 Z\"urich, Switzerland}  \\[4pt]
  {\em $(b)$  Dipartimento di Fisica e Astronomia ``G. Galilei'', \\ Universit\`a di Padova, Via Marzolo 8,
I-35131 Padua, Italy }

\end{center}
\vspace*{1.0cm}

\centerline{\large\bf Abstract}
\medskip\noindent 
We analyse the predictions of both Higgs and top masses
in a generic MSSM satisfying gauge-coupling unification, radiative electroweak symmetry breaking, 
with a natural (non-splitted) spectrum of soft-breaking terms, and an arbitrary soft-breaking scale (above a few TeV). 
This procedure allows us  to identify a relatively narrow SUSY-allowed region in the $(m_h, m_t)$ plane.
We argue that the compatibility of the measured values of Higgs and top masses with this  SUSY-allowed region is not less 
surprising than the corresponding compatibility with the narrow 
SM metastability region. As such, it provides 
a non-trivial signal of the compatibility of present data with a supersymmetric completion of the SM.

\tableofcontents
\newpage

\section{Introduction}

Among the possible ultraviolet completions of the Standard Model (SM),
supersymmetric  theories represent one of the most motivated and attractive options.
This statement remains true despite the absence, so far reported by LHC experiments,  
of direct signals of new particles above the electroweak scale. 
These negative experimental results  
 indicate that low-energy supersymmetry (SUSY) 
cannot provide a completely natural solution to the hierarchy problem of the electroweak scale.
However, many additional potential virtues of supersymmetric  theories, 
such as gauge coupling unification, the stabilisation of the Higgs potential, 
a natural dark-matter candidate, and the attractive possibility of coherently 
embedding  gravitational interactions in the context of supergravity, remain open.

Exploring supersymmetric extensions of the SM
abandoning the criteria of a natural solution to the hierarchy problem 
is a rather difficult task. 
Even more difficult, but somehow more interesting, is trying to address the question of how likely is to reproduce 
the observed SM spectrum, and the absence of new-physics signals, 
in a given supersymmetric model. 
The purpose of this paper is  to provide a partial answer to the latter question,
in the context of the Minimal Supersymmetric extension of the Standard Model (MSSM). 

Abandoning the naturalness criteria, significant constraints on the MSSM 
parameter space can be derived exploiting the renormalisation group (RG) flow of the theory,
or the connection between low- and high-energy scales. 
In this context, the value of the Higgs mass ($m_h$)  and the requirement of 
gauge coupling unification provides two key ingredients.  
The former can be viewed as a prediction of the RG flow in the infrared, while the 
latter can be used to anchor the parameter space of the theory in the ultraviolet.

These two features of the MSSM have been extensively discussed in the past 
and recent literature to derive upper or lower bounds on the SUSY breaking scale 
(see e.g.~Ref.~\cite{Ellis:1990wk,Amaldi:1991cn,Langacker:1991an,Ellis:1990nz,Haber:1993an,Carena:1995wu,Heinemeyer:1998yj,ArkaniHamed:2004fb,Giudice:2004tc,Hall:2009nd,Giudice:2011cg,Arvanitaki:2012ps,Bagnaschi:2014rsa,Vega:2015fna,Ellis:2017erg}).
As is well know, $m_h$ is predicted to lie below the $Z$--boson mass in the MSSM at tree level.
This relation receives quantum corrections that grow logarithmically with 
the scale of the soft-breaking terms ($m_S$). As a result, the experimentally measured value, $m_h=125.1 \pm 0.2$~GeV,
provides a significant lower-bound on $m_S$. The latter has been investigated in great detail after the Higgs boson
discovery~\cite{Giudice:2011cg,Arvanitaki:2012ps,Bagnaschi:2014rsa,Vega:2015fna}.
Conversely, the requirement of gauge-coupling unification
can be used to derive an upper bound on $m_S$.  Barring the case of a huge mass splitting among the soft-breking terms, as in 
split-SUSY~\cite{ArkaniHamed:2004fb,Giudice:2004tc}, 
gauge coupling unification can be achieved only if $\mS$ is low enough to correct the non-unifying RG flow of the SM.
According to the recent analysis of~\Ref\cite{Bagnaschi:2014rsa}, this occurs for $\mS$ up to about $10^6$ GeV.

The purpose of this work is to further explore these and other features of the RG flow of the MSSM, 
with a purpose different than in most previous analyses, namely trying to address the question of how likely 
is to reproduce the observed SM spectrum (or better some of its key features). More precisely, 
our concrete goal is to investigate the predictions of both Higgs and top ($\mt$) masses
in a generic MSSM, satisfying gauge-coupling unification and radiative electroweak symmetry breaking, 
with a non-splitted spectrum of soft-breaking terms. 

The reason to focus the attention on both $\mh$ and $\mt$ is twofold.
On the one hand, these are the masses of the two heaviest SM fields. 
On the other hand, these low-energy observables are in one-to-one correspondence 
with the Higgs self-coupling and the top Yukawa coupling, namely the two leading non-gauge 
interactions of the theory. The combined values of $\mh$ and $\mt$, together with the Fermi coupling, 
thus provide a clear characterisation of the symmetry breaking sector of the theory at low energies. 
This statement acquire a special meaning in the context of the MSSM, where the 
top Yukawa coupling may trigger a radiative breaking of the electroweak symmetry,
starting from a symmetric configuration at high scales. 

Investigating the MSSM predictions in the $(\mh,\mt)$  plane is also of particular interest 
when comparing them with those obtained in the non-SUSY case, extrapolating the 
validity fo the SM up to the Planck scale. The latter analysis has triggered a lot of attention 
in the last few years given the peculiar position of the measured  $\{\mh^{\rm exp},\mt^{\rm exp}\}$ point,
which lies in the narrow SM metastability region~\cite{Degrassi:2012ry,Bezrukov:2012sa,Buttazzo:2013uya}. 
As we will argue in the following, the interpretation of the same 
point in a landscape of  MSSM models is not less interesting, and can be interpreted as 
a non-trivial signal of the compatibility of present data with a supersymmetric completion of the SM.

\section{Assumptions and strategy of the analysis}  \label{secF_method}

As anticipated in the introduction, exploring the MSSM parameter space --- abandoning the criteria of a natural solution to the hierarchy problem as main guide to restrict it --- 
is a non-trivial task.
In order to achieve this goal, we need additional hypotheses.
Our strategy is based on the following four main requirements:
\begin{itemize}
			\item  gauge coupling unification;
			\item  natural range for the soft-breaking  terms at the GUT scale (natural soft spectrum);
			\item  radiative breaking of the electroweak symmetry (REWSB);
			\item  prediction of the Higgs mass ($\lambda$-matching).
\end{itemize}
The first two conditions bound the structure of the model at high energies, 
while the last two conditions are related to the matching into a SM-like theory at low energies. 
In the following we first formulate in a more precise way these requirements, translating them 
into well-defined quantitative constraints on the soft-breaking terms. 
Next we illustrate the strategy adopted to  efficiently span  the MSSM parameter space
checking if and where these conditions are satisfied. 

\subsection{Quantitative definition of the requirements}
\label{ssecF_b_def}

\begin{itemize}
\item{} {\em Gauge coupling unification.} In order to implement the condition of gauge unification, 
we need to quantify the amount of threshold corrections we are ready to tollerate at high scales. 
This issue has  been extensively studied in the literature and will not be reanalysed here in detail. 
We will simply exploit the results recently obtained in  \Ref\cite{Bagnaschi:2014rsa}. According to this analysis, 
a satisfactory gauge-coupling unification for a non-splitted spectrum (defined as in the
item below) does occur if $\mS \lesssim 10^6$ GeV.

Taken for granted that gauge coupling unification is achieved, we should only care about a sensible definition of the GUT scale itself, $\mG$.
Given that a small residual mismatch among gauge couplings always remain, it is reasonable to define $\mG$ in such a way to minimise it.
For this reason, we will operatively define $\mG$ as:
\begin{align}			
			\mG = \min_{\mu} \left(
						\frac{\sqrt{(g_1(\mu) - g_2(\mu))^2 + (g_2(\mu) - g_3(\mu))^2 + (g_1(\mu) - g_3(\mu))^2}}
						{g_1(\mu)+g_2(\mu)+g_3(\mu)} \right)
			\pun
			\label{eqF_b_GUTdef}
\end{align}
A different choice, common in the literature, could have been to simply set $\mG$ as the energy at which $g_1(\mu)=g_2(\mu)$.
It is worth noting that different definitions may lead to $\order1$ variations in the exact value of $\mG$.
However, being the sensitivity of the RG flow to the input scale only logarithmic, it is intrinsically irrelevant for our purposes to define $\mG$ 
with an accuracy better than $\order1$.

\item{} {\em Natural soft spectrum.}
The definition of the acceptable range for the variation of the the soft breaking terms
(and of the $\mu$ parameter)  is one of the 
key issue of the present analysis. 
First of all, we stress that we define this range at the high scale $\mG$, where such terms are presumably generated. 
At this scale we consider a configuration of (flavour-blind) soft breaking terms as natural (hence acceptable in our analysis) if the ratios of independent terms with respect to a common input scale lies within $1/r$ and $r$, with $r=O(1)$.
Note that this condition is qualitatively different from the requirement of minimal fine-tuning in deriving a low-energy 
observable (such as the Higgs mass or the electroweak scale) and it follows only from the physical assumption 
of a common origin of the soft-breaking terms. 

Operatively, the common scale is conventionally chosen to be the gluino mass, $M_3(\mG)$, and 
we set $r=3$, although the implications of larger values of $r$ are also investigated.  
More precisely, dimension-two soft couplings (sfermion and Higgs masses) 
are varied in the natural range in units of $M^2_3(\mG)$, while
dimension-one soft couplings  are varied in units of $M_3(\mG)$. 
To simplify the analysis, we further impose the relation $M_1(\mG) = M_2(\mG) = M_3(\mG)$
among the gaugino masses. 

\item{} {\em Radiative electroweak symmetry breaking.}
The first low-energy condition we impose on the model in order to reduce the number of independent 
parameters, or better to express them in terms of low-energy observables, 
is the requirement of electroweak symmetry breaking. This imply the two relations
\begin{align}
			\left\{ \ba{l}
			\mHu^2 + |\mu |^2 - b \cot\beta - \frac12  \, m_Z^2 \cos (2\beta) = 0	\phantom{\Big(}	\\
			\mHd^2 + |\mu |^2 - b \tan\beta + \frac12 \, m_Z^2 \cos (2\beta) = 0	\phantom{\Big(}
			\ea \right.
			\label{eqE_b_EWSBmatch}
\end{align}
where 
\be
 v_{u(d)} = \langle H_{u(d)}^0\rangle~,  \qquad \tan\beta \equiv \frac{v_u}{v_d}~, \qquad 
 v_u^2 + v_d^2 = v^2 \approx (174 \, {\rm GeV})^2~,
\ee
which can be solved only if
\begin{align}
			(|\mu|^2 + m^2_{H_u} )(|\mu|^2 + m^2_{H_d}) < b^2~.	
			\label{eqE_b_vacReq2}
\end{align}
This relation allows us to express some of the soft-breaking terms and $\mu$ in terms of
 $m_Z$ and $\tan\beta$. Out of these two low-energy  observables, 
 only  $m_Z$ is  fixed to its physical value (i.e.~is considered a fixed condition in our parameter scan).
 
As anticipated, we are interested in studying under which conditions the mechanism of electroweak 
symmetry breaking occurs as a radiative effect at low energies, starting from a symmetric condition at
high scale (i.e.~$m_G$).
This occurs if \eq(\ref{eqE_b_vacReq2}) is not fulfilled at $\mG$,  namely 
if 
\begin{align}
			\left( |\mu(\mG) |^2 + m^2_{H_u}(\mG) \right)
			\left( |\mu(\mG) |^2 + m^2_{H_d}(\mG) \right) > b^2(\mG)
			\pun
			\label{eqF_b_REWSB}
\end{align}
The simultaneous requirement of \eq(\ref{eqE_b_vacReq2}), at the electroweak scale, and \eq(\ref{eqF_b_REWSB}),
will be our operative definition of REWSB.

\item{} {\em Higgs mass.}
The second low-energy constraint on the model is derived by the matching  of the effective 
Higgs quartic coupling, $\lambda_\sSM$, evolved up the scale $\mS$ according the SM RG flow (starting from the electroweak scale),
with its prediction in terms of MSSM parameters, $\lambda_\MSSM(\mS)$, as sketched here:
\begin{align}						
\lambda_\MSSM(\mS)	&= \frac14 \left[ g^2(\mS) + g^{\prime 2}(\mS) \right] \cos^2 2\beta+ \Delta \lambda^{1 \ell  oop}~.
\label{eqF_a_lamMatch}
\end{align}
Here $\Delta \lambda^{1 \ell  oop}$ denotes the complete one-loop correction to the tree-level matching condition, 
whose explicit expression can be found e.g.~in Ref.~\cite{Giudice:2011cg,Bagnaschi:2014rsa}. 
Since $\Delta \lambda^{1 \ell  oop}$ depends on the sparticle spectrum and other MSSM parameters at the scale $\mS$,
\eq(\ref{eqF_a_lamMatch}) provide a well-defined low-energy matching condition.
From now on we will refer to it as the $\lambda$-matching condition.

Ideally, we should impose $\lambda_\sSM(\mS)=\lambda_\MSSM(\mS)$; however, since both the RG equations 
and the matching expression are derived in perturbation theory, the neglect higher-order terms induced an unavoidable mismatch.
Therefore, we impose the $\lambda$-matching condition in the following form 
\begin{align}
			|\lambda_\sSM(\mS) - \lambda_\MSSM(\mS) | < \ltr \sim \max \left( \Delta\lambda^{2\ell oop} \right)~,
			\label{eqF_b_lamMatch}
\end{align}
where $\ltr$ generically denotes the maximal amount of (low-energy) threshold corrections we are ready to tollerate. 
As indicated in \eq(\ref{eqF_b_lamMatch}), the latter are estimated using the  leading two-loop 
threshold corrections ($\Delta\lambda^{2\ell oop}$).
The precise value of  $\ltr$, that we vary as a function of  $\mS$, is reported in Appendix~\ref{sect:appA}.
\end{itemize}

\subsection{Analysis strategy} \label{ssecF_b_glob}

In this section we illustrate the procedure followed to analyse the MSSM parameter space taking into account 
the four main requirements specified above.
Let us recall that we want to perform our study considering different input values of $\mh$ and $\mt$,
keeping the Fermi scale and the SM gauge couplings (hence $W$ and $Z$ masses) fixed
to their physical values.
For convenience, the Fermi scale is expressed in terms of $\mtp = (2\sqrt{2} G_F)^{-1/2}  \approx 174$~GeV.
Namely  the Higgs self-coupling and the top Yukawa coupling are determined 
in terms of the physical values of $\mh$ and $\mt$ at $\mu=\mtp$.

\begin{table}[t]
			\centering
			\begin{tabular}{| c | c c c c c   |}
			\hline
			\interlineB
			{\small Step} &
			$\mtp$ &   & $\mS$ &   & $\mG$ 			\\
			\doubleinterline
			\interlineB
			1 &		$\gA(\mtp)$ & $\stackrel{\rm SM-RGE}{\longrightarrow}$ &
						$\gA(\mS)$ & $\stackrel{\rm MSSM-RGE}{\longrightarrow}$ & 
						$\gA(\mu) \to \mG$ [\eq(\ref{eqF_b_GUTdef})]  	\\
			\interline
  			2 &		&& $M_3(\mS) = \mS$  
  						&  $\stackrel{\rm MSSM-RGE}{\longrightarrow}$ & $M_3(\mG)$ 	\\
					\interlineB
 						 & $  \ba{c}  m_t \to \gt(\mtp)  \\[3pt] 
						             \mh \to \lambda(\mtp)  \ea $ & $\stackrel{\rm SM-RGE}{\longrightarrow}$
 						 &  $  \ba{c}  \yt(\mS)  \\[3pt] 
						             \lambda(\mS)  \ea $  && \\
			\interline
			3 	&&& $\left\{ \begin{array}{l}
				 					M_{1,2}(\mS) \\ \At(\mS) \\ \mQU(\mS) \\ \mHud(\mS)
				 					\end{array} \right. $
						& $\stackrel{\rm MSSM-RGE}{\longleftarrow}$ 
						& $\left\{\ba{l}  M_{1,2}(\mG) \\ \At(\mG) \\ \mQU(\mG) \\ \mHud(\mG) \ea \right.\!$ 
						 [\eq(\ref{eqF_b_gutBCs})]  \hfill
						 \\
					\interlineB
						&&& $\downarrow$  &     &     \\
					\interlineB
						&&   \hfill  [\eq(\ref{eqE_b_EWSBmatch})]   & $\left\{ \ba{c} \mu(\mS) \\ b(\mS) \ea \right. \hfill$ 
						& $\stackrel{\rm MSSM-RGE}{\longrightarrow}$ 
						& $\left\{ \ba{c}  \mu(\mG)  \\ b(\mG)   \ea \right. \hfill$      \\ 
						\interlineB 
						&&  \hfill check:  &  $\ba{l} \bullet~\text{EWSB} \\
										\bullet~\lambda-\text{matching\!\!\!\!\!\!\!} \ea$
					        &&  check:   $\ba{l} \\ \bullet~\text{REWSB} \\
										\bullet~\text{Natural}  \\ 
										\phantom{\bullet}~\text{soft~spectrum}\ea$  \hfill    \\ 
			\interlineB
			\hline
			\interlineB
			4 & 	\multicolumn{5}{l |}
						{Characterisation of  the $( \mh,\mt )$ point according to Step 3 results} \\
			\hline 
			\end{tabular}
			\caption{Schematic representation of the analysis performed in order to characterise each point in the 
			$( \mh,\mt )$ plane.
			\label{tabF_steps}}
\end{table}
%

The procedure adopted is based on four steps, schematically illustrated in 
Table \ref{tabF_steps}, which can be summarised as follows.
\begin{itemize}
			\item Step 1: determination of the GUT scale  [{\em input}~: $\mS$]. \\
						For a given $\mS$ value,  the SM gauge couplings (whose low-energy values are taken from experiments) 
						are evolved up to $\mS$
						using SM one-loop RGE equations,  and further up using supersymmetric RG equations.
						The GUT scale $\mG$ is then determined by means of \eq(\ref{eqF_b_GUTdef}).

			\item Step 2: determination of boundary conditions  at $\mS$ and $\mG$ [{\em input}~: $\{ \mh, \mt, \tan\beta\} $]. \\ 
						Starting from the chosen values of $\mh$ and $\mt$ (denoting the corresponding pole masses)
						the values of the Higgs coupling $\lambda$  and the 
						and top Yukawa coupling $\gt$, within the SM, are determined (by means of one-loop matching conditions) 
						at the electroweak scale $\mtp$ and evolved up to $\mS$. 
						For a given $\tan\beta$ value, the supersymmetric top Yukawa  coupling ($\yt$) is fixed by the (tree-level)
						relation $\yt(\mS)=\gt (\mS) \sin\bS$. \\
						Identifying the gluino mass with the soft-breaking scale $M_3(\mS)=\mS$, and evolving it
						up to the GUT scale, we determine $M_3(\mG)$. As anticipated, the latter is 
						used as reference  scale to define the natural range 
						of all the other soft-breaking parameters  at the GUT scale.

			\item Step 3:  scan over the remaining MSSM parameters. \\
						At the end of step 2 only the gaugino masses 
						and the top Yukawa coupling have been determined. 
						The parameters still unknown and relevant for our analysis are $\{\mQU, \At, \mHud, \mu, b\}$.
						However, not all of them are independent since we are interested only in configurations giving rise to electroweak
						symmetry breaking  (with the physical value of the Fermi scale). \\
						To scan the viable parameter configurations we use the following strategy: 
						we fix $\{\mQU, \At, \mHud \}$ at the GUT scale following our criterium of a natural soft-breaking spectrum, namely we vary 						
						them in the range
						\begin{align}
								\left\{ \begin{aligned}
										\mHud^2(\mG)		& 	= r_{H_{u,d}} \, M_3^2(\mG) \\
										\mQU^2(\mG)		& = r_{Q,U} \, M_3^2(\mG) \\
										\At(\mG) 				& = \rt \, M_3(\mG)
								\end{aligned} \right.~,
								\qquad
								\rA \in \left( \frac{1}{\bar r} , \bar r \right) \vir
								\label{eqF_b_gutBCs}
						\end{align}
						with $\bar r = 3$. We then run them down to $\mS$ and invert \eq(\ref{eqE_b_EWSBmatch}) to determine $|\mu(\mS)|$ and $b(\mS)$.
						Having obtained the full soft-breaking spectrum at the low scale we are able to determine 
						$\lambda_\MSSM(\mS)$ and check if the $\lambda$-matching condition is satisfied. 						
						The low-energy values of  $|\mu(\mS)|$ and $b(\mS)$ are then evolved up to $\mG$ in order to check 
						if they are compatible with the assumption of a natural spectrum,
						i.e.~if 
									\begin{align}
											\rmb \in \left( \frac{1}{\bar r} , \bar r \right)~, \qquad \mbox{where} \qquad 
														\quad
														\left\{ \begin{aligned}
																\rmu  &\equiv \mu(\mG)^2 / M_3(\mG)^2 	\\
																\rb  &\equiv b(\mG) / M_3(\mG)^2 
														\end{aligned} \right.~,
														\label{eqF_b_mubNat}
									\end{align}
						and  to check if the electroweak symmetry breaking has been achieved radiatively. \\
						Summarising, for each configuration of soft-breaking terms we check the following conditions:
						\begin{itemize}
									\item Consistent electroweak symmetry breakdown ($|\mu|^2 >0$); 
									\item The $\lambda$-matching condition in \eq(\ref{eqF_b_lamMatch});
									\item The naturalness of $\mu(\mG)$ and $b(\mG)$ defined in Eq.~(\ref{eqF_b_gutBCs});
									\item The REWSB  condition, \eq(\ref{eqF_b_REWSB}).
						 \end{itemize}
						
			\item Step 4: characterisation of the $(\mh,\mt)$ point. \\
					     The outcome of the scan in step 3 (including also a scan over the value of $\tan\beta$   in the range $\tan\bS  \in [1.1, 10]$ from step 2) 
					     is classified according to the maximal number of conditions satisfied by at least 
					     one configuration of soft-breaking terms. In particular, we classify the point in the 
					     $(\mh,\mt)$ plane as a {\em viable point} if at least the first three conditions (i.e.~all but REWSB) are satisfied.
\end{itemize}

As it can easily be understood,  the most challenging task from the 
computational point of view is the scan over $\{\At, \mQU^2, \mHud^2\}$ in step 3.
We find that a reliable scan of such five-dimensional parameter space requires to probe 
$\order{10^4}$ independent configurations for each step 2 point.
Since a full numerical handling of such RG flow would be prohibitively time consuming, 
whenever possible analytical solutions to the RG equations of the soft parameters have been developed and implemented.

\begin{figure}[t]
			\centering
			\includegraphics[width=0.4 \textwidth]{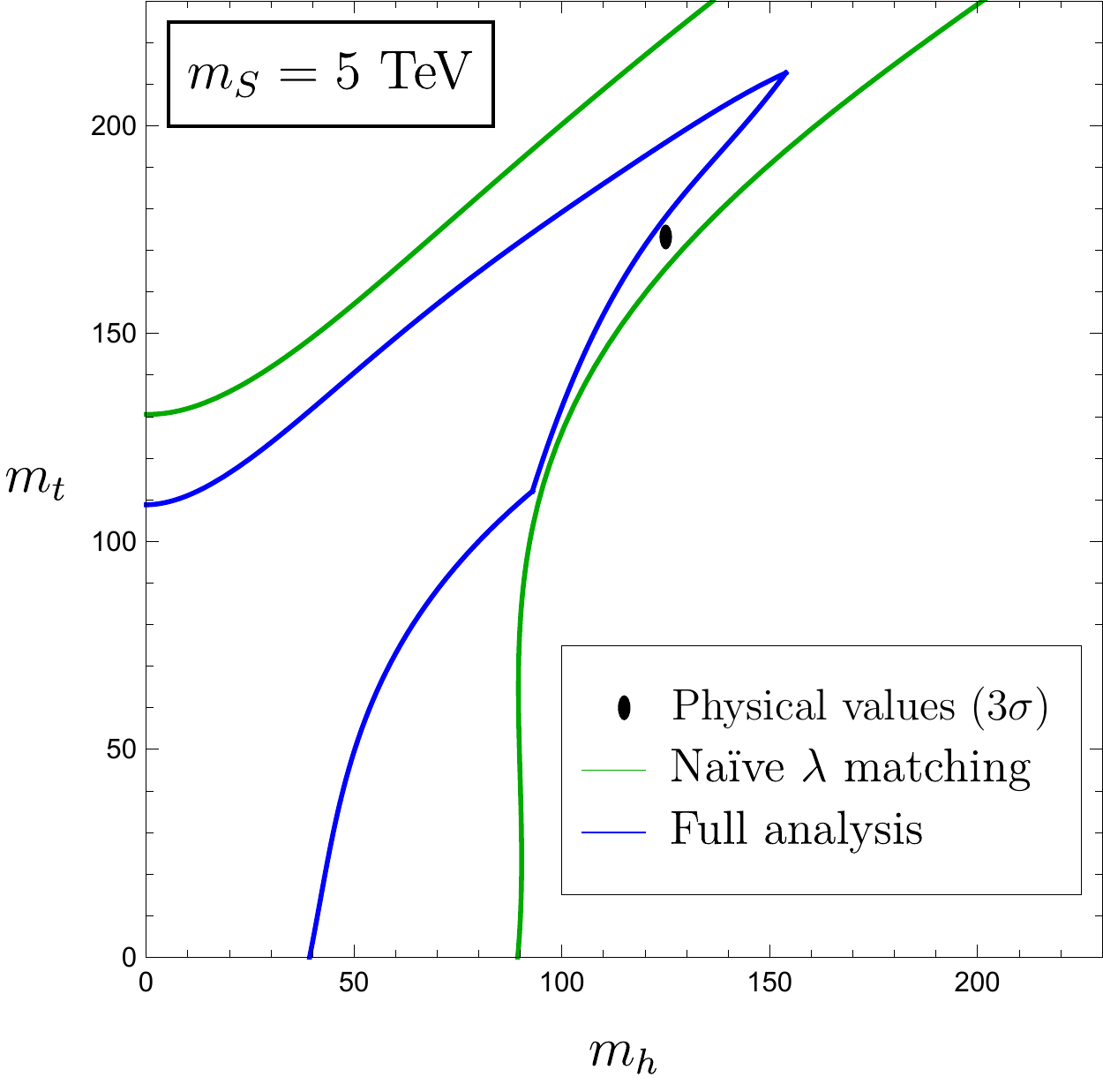}
			\caption{Comparison of the 
			viable region in the  $(\mh,\mt)$ plane obtained according to our analysis, without imposing 
			the REWSB condition  (blue line), and the one following only from a na\"ive matching condition on the Higgs mass (green line). 
			The physical values for $(\mh,\mt)$ are also shown for reference.
			\label{figF_5TeV_BG}}
\end{figure}

\section{Discussion of the results}
\label{secF_res}

\subsection{Analysis of the constraints in the $(\mh,\mt)$ plane}

We are now ready to discuss the viable region in the  $(\mh,\mt)$ plane identified 
following the analysis illustrated in Sect.~\ref{secF_method}.
We first discuss the results obtained at low $m_S$, without imposing the requirement of REWSB.
Wen then proceed discussing the impact of the latter requirement and of increasing $m_S$ values.

\subsubsection{The low $m_S$ scenario}

In Figure \ref{figF_5TeV_BG} we show the viable $(\mh,\mt)$ region for $\mS=5~\TeV$. More explicitly, we compare the results obtained 
according to our analysis, without the REWSB requirement (blue line, {\em full analysis}), with those obtained using only the 
 $\lambda$-matching condition (green line, {\em  na\"ive $\lambda$-matching}). The latter is defined as the region obtained 
 imposing only Eq.~(\ref{eqF_b_lamMatch}), scanning over the relevant MSSM parameters directly at the matching scale (no RG effects above $\mS$), 
 maximising in size the finite part of the one-loop threshold corrections (with positive or negative sign depending on the border).

The two boundaries of the na\"ive viable region can be understood as follows: i) for points on the right of the lower green curve, the quartic Higgs coupling is too high 
to be properly matched with its MSSM expression in terms of SM gauge couplings (even when maximal threshold corrections are taken into account); ii) 
for points above the upper green curve, $\lambda_\sSM(\mS)$ 
becomes negative (of an amount that cannot be accommodated through threshold corrections).

Our viable region (in blue) is obviously more restrictive being based on more demanding conditions. 
Its peculiar shape can be divided into three different pieces: an upper border (representing the highest allowed $\mt$ value for a given $\mh$), 
a right border (that goes from the end of the up border down to a visible ``shoulder'') and a lower border (that from the above mentioned shoulder goes down to $\mt=0$).
Each of these curves has a different origin and explanation.

\FPeval{\nA}{0.4}
\begin{figure}[t]
			\centering
			\begin{minipage}{\nA\textwidth}
			\centering
			\includegraphics[width=0.9\textwidth]{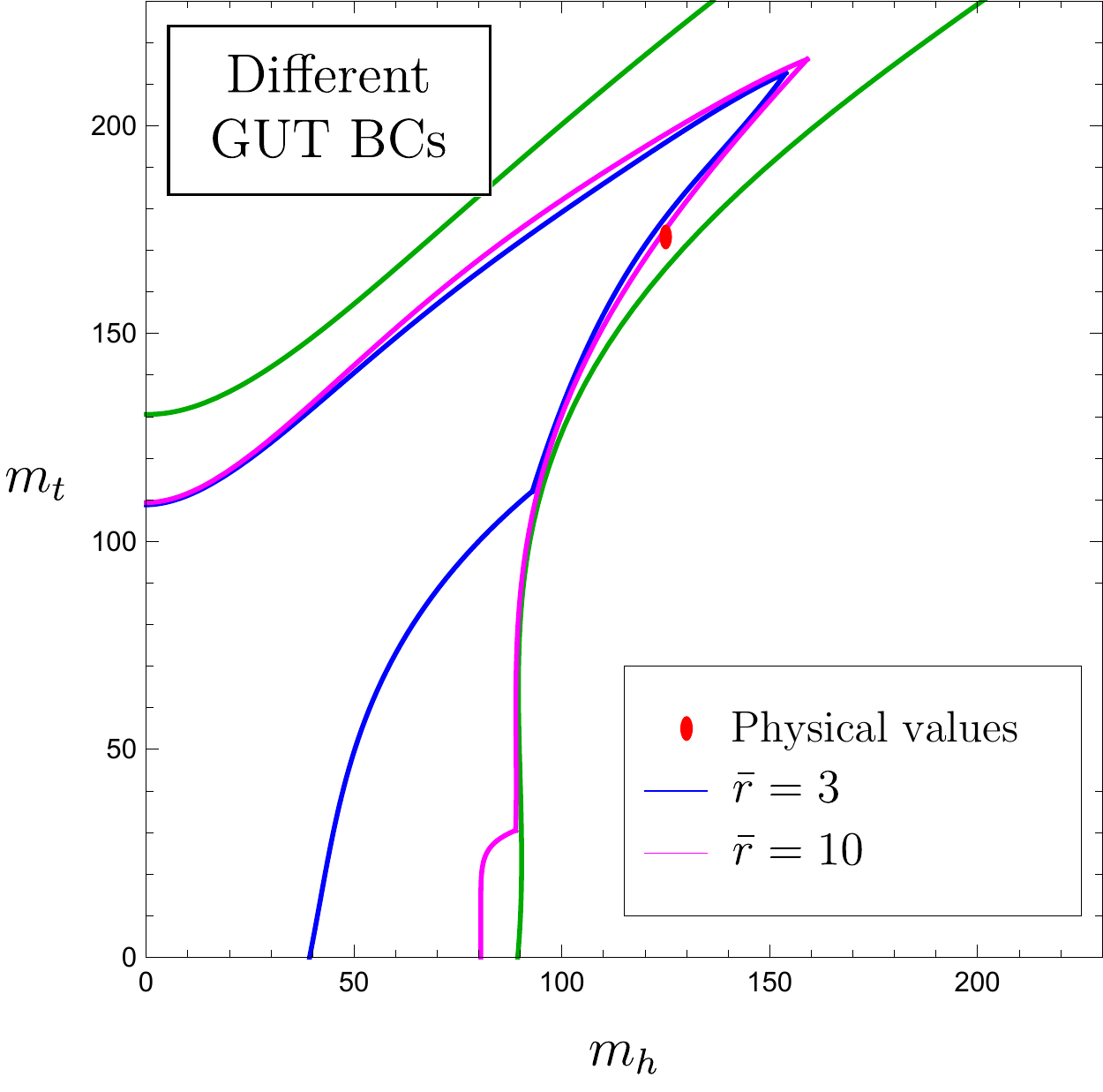}
			\end{minipage}
			\begin{minipage}{\nA\textwidth}
			\centering
			\includegraphics[width=0.9\textwidth]{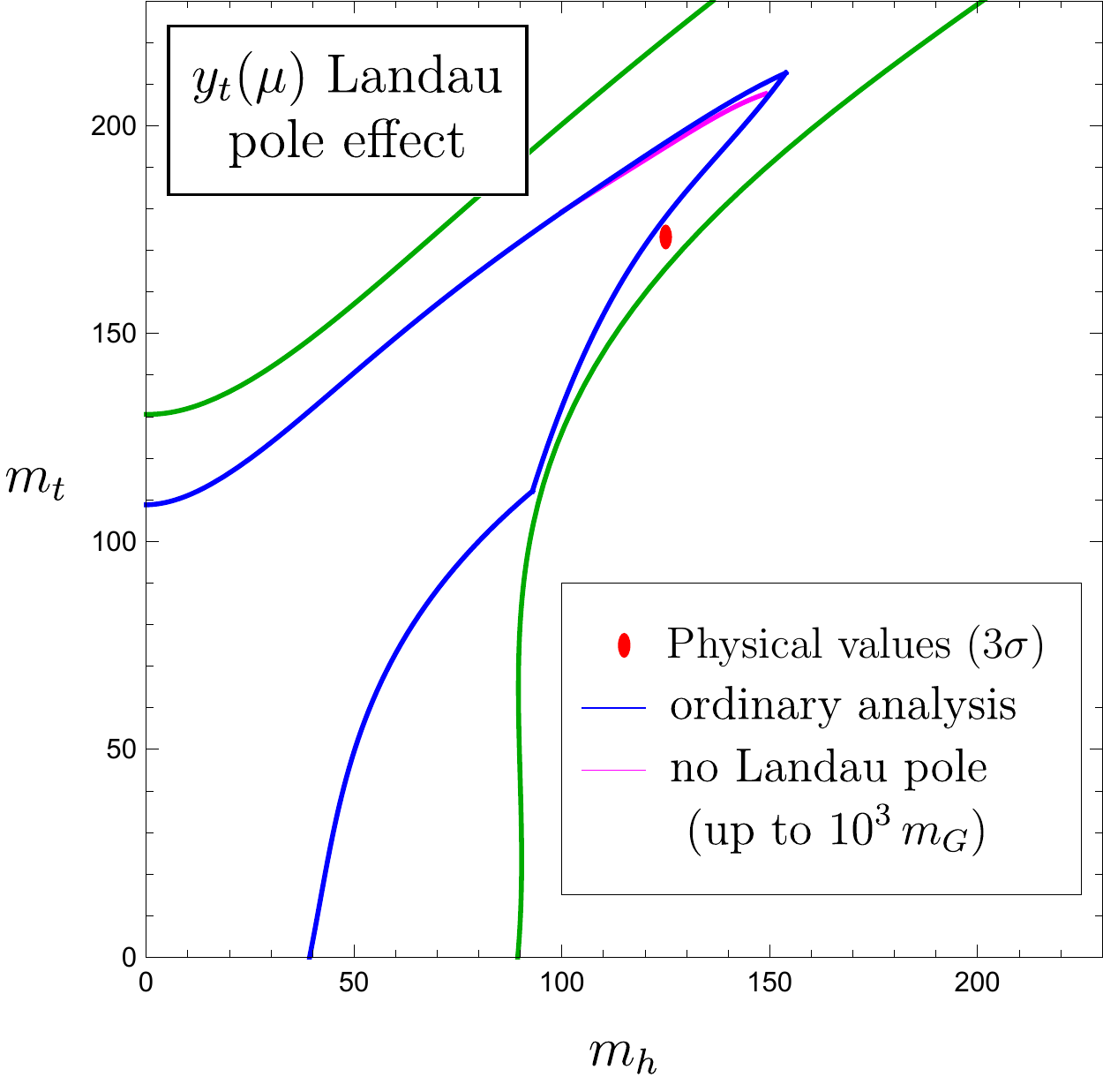}
			\end{minipage}
			\begin{minipage}{\nA\textwidth}
			\centering
			\includegraphics[width=0.9\textwidth]{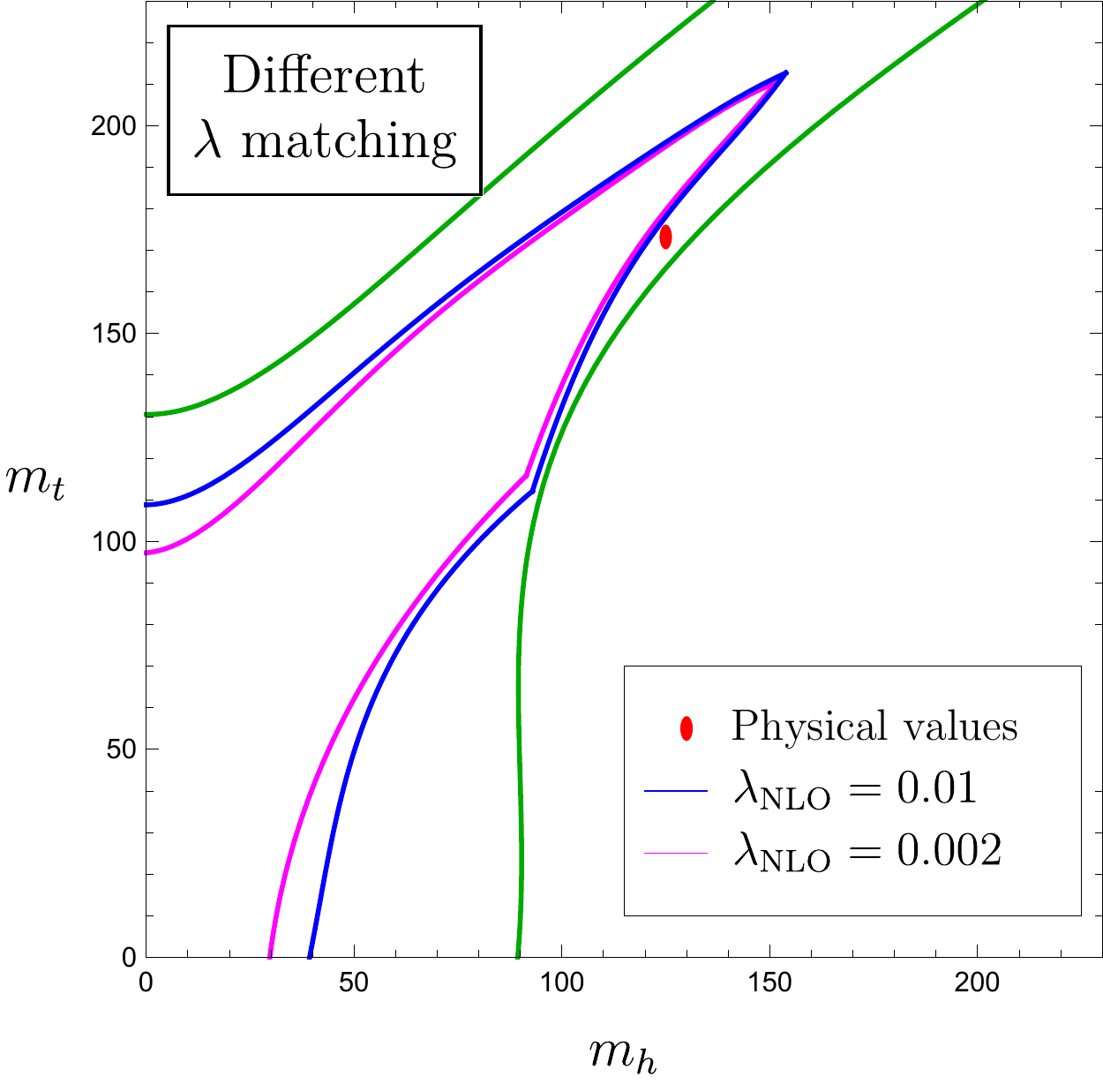}
			\end{minipage}
			\begin{minipage}{\nA\textwidth}
			\centering
			\includegraphics[width=0.9\textwidth]{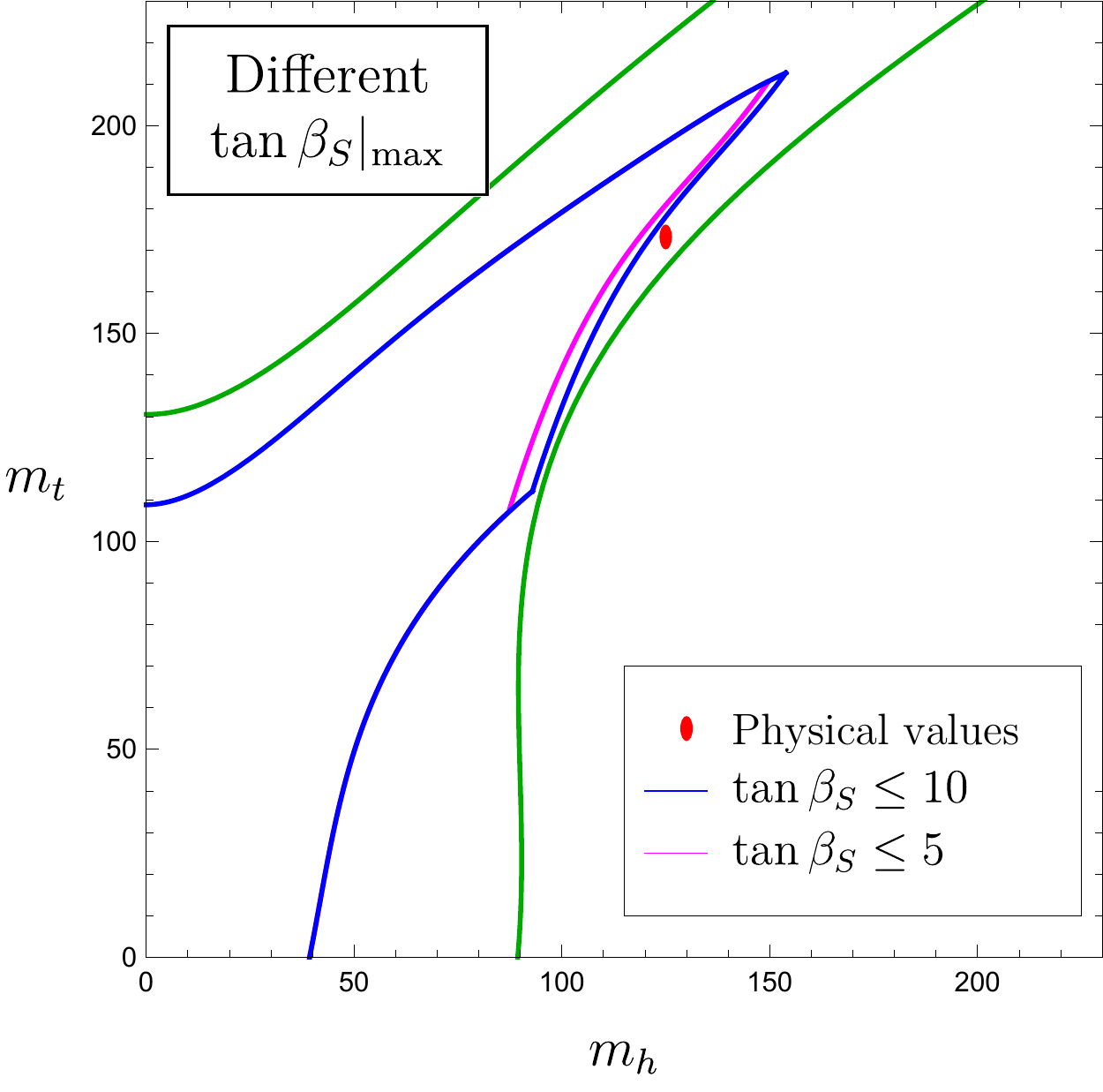}
			\end{minipage}
			\caption{Comparison between the viable region for the reference set of parameters (in blue) and the
								regions resulting from modifications of some of these parameters (in magenta, see 
						            text for more details). The green curves of Figure \ref{figF_5TeV_BG} are also shown for
								reference. 
			\label{figF_5TeV_comparison} }
\end{figure}

\begin{itemize}
			\item The upper border has the same origin as the (green) upper border from the na\"ive $\lambda$-matching: 
			         it denotes the region above which $\lambda_\sSM(\mS)$ becomes too negative 
				 to be rescued by threshold corrections. 
				 The blue curve is lower that the green one since in the full analysis threshold corrections are limited by the assumption 
				 of a natural spectrum at the GUT scale. Moreover, in the high-$\mt$ part of the border (i.e.~for $\mt \gsim 190$~GeV),
				 an additional effect  enters the game: the requirement that $\yt(\mu)$ does not develop a Landau pole before reaching the
				GUT scale. This forces us to raise $\tan\bS$ as $\mt$ increases (since this lowers, in turn, the low-scale value of $\yt$), 
				correspondingly affecting the $\lambda$-matching condition.
				
			\item Similarly, the right border has the same origin as the corresponding right border from the na\"ive $\lambda$-matching: 
				it denotes the region after which $\lambda_\sSM(\mS)$ becomes too large.
				As for the upper border, the shift between green and blue curves is due to the limitation on the threshold 
				corrections.  Indeed the two curves tends to overlap at low $\mt$, where threshold corrections 
				plays a smaller role. 

			\item The lower border  has no analog in the the na\"ive $\lambda$-matching case:
				for $\mt$ values below this curve, the smaller value of $\tan\bS$ required by 
				$\lambda$-matching does not allow us to fulfil  the  electroweak symmetry breaking condition
				(within our boundary conditions on the soft spectrum at the GUT scale). 
				On this effect, two comments are in order:
						\begin{itemize}
									\item For low $\mt$, the tension between $\lambda$-matching and EWSB, \eq(\ref{eqE_b_EWSBmatch}), is clear:
												the former prefers high values for $\tan\bS$, in order to increase 
												$\lambda_\MSSM$;
												the latter prefers low values for $\tan\bS$ to compensate for the small $\mt$ value,
												which reduces the $\yt$ influence on the RG flow.
									\item In the absence of big RG 
												flow effects, the achievement of electroweak symmetry breakdown 
												depends critically on the choice of boundary conditions imposed 
												at $\mG$.  In particular, the excluded region tends to shrink 
												for larger values of $\bar r$ in \eq(\ref{eqF_b_gutBCs}).
								\end{itemize}
\end{itemize}

In order to better understand the features discussed  above, it is interesting to study how the various boundaries change by varying the parameters of the analysis. 
An illustration of the main effects is shown Figure \ref{figF_5TeV_comparison}.

\begin{figure}[t]
			\centering
			\includegraphics[width=0.4 \textwidth]{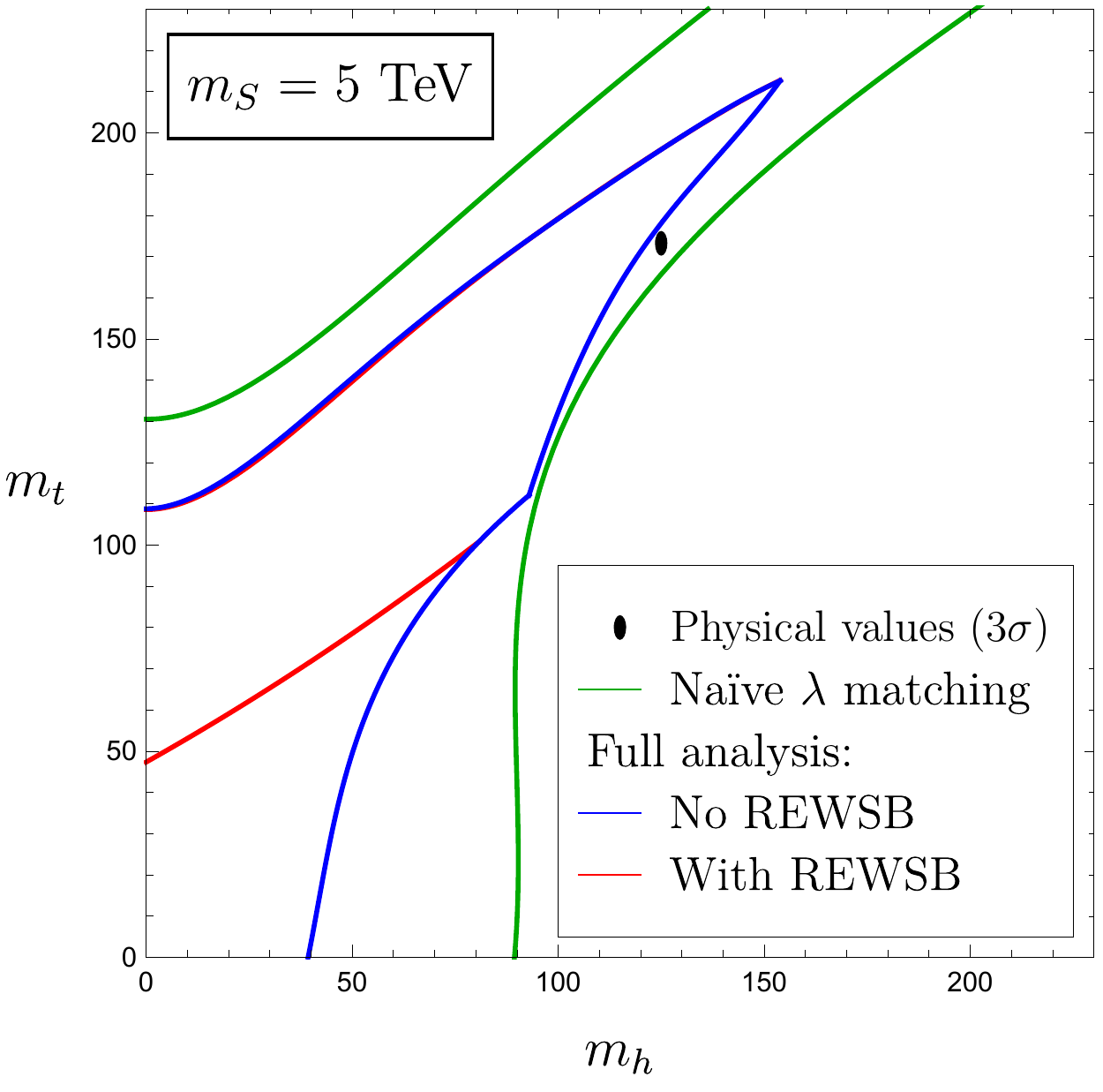}
			\caption{Na\"ive viable region (green) and viable region with (red) and without (blue) the additional
								REWSB condition. 
						\label{figF_5TeV_BRG}}
\end{figure}
\begin{figure}[t]
			\centering
			\mbox{
			\begin{minipage}{0.45 \textwidth}
						\centering
						\includegraphics[width=0.9\textwidth]{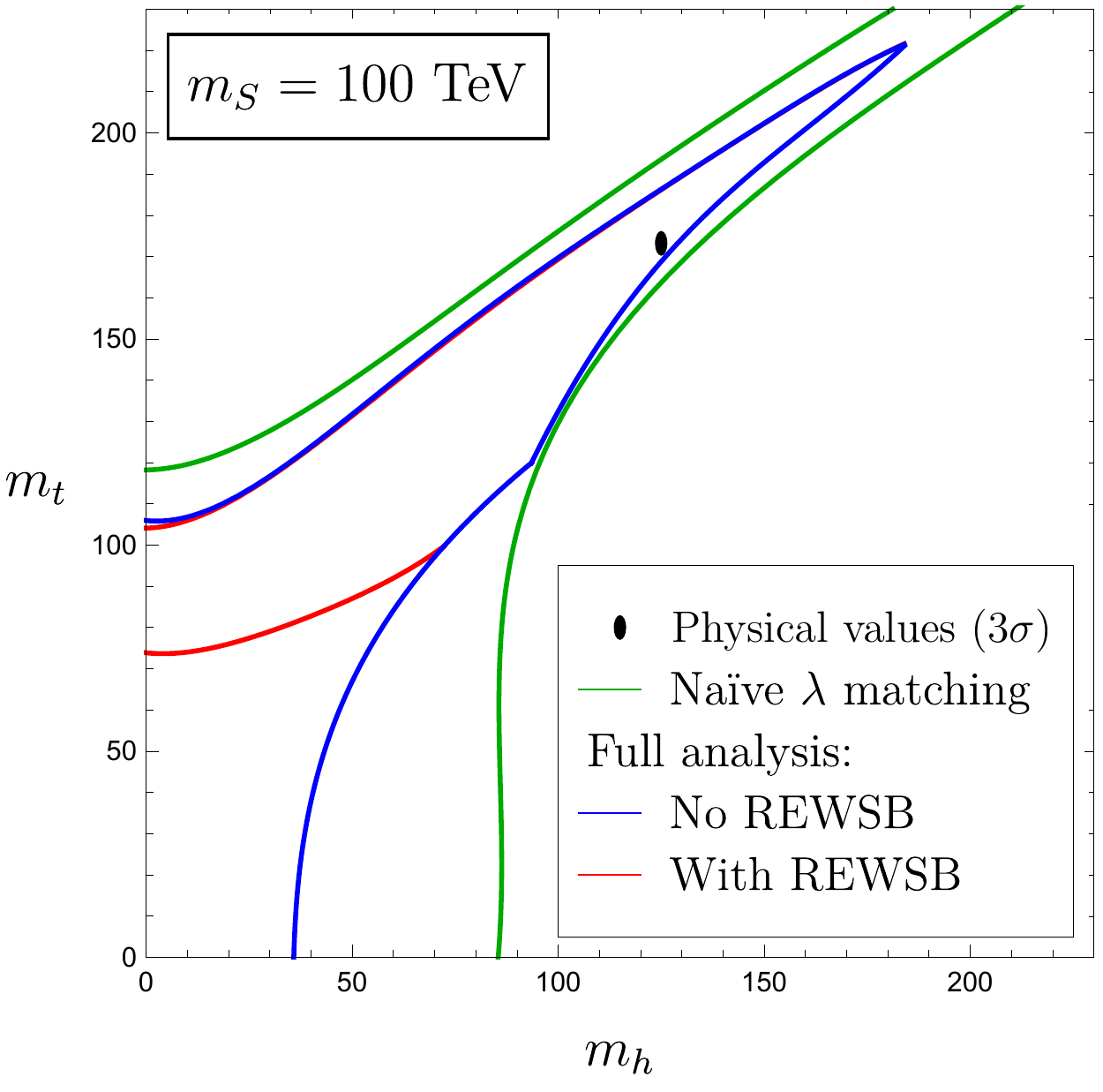}
			\end{minipage}
			\begin{minipage}{0.45 \textwidth}
						\centering
						\includegraphics[width=0.9\textwidth]{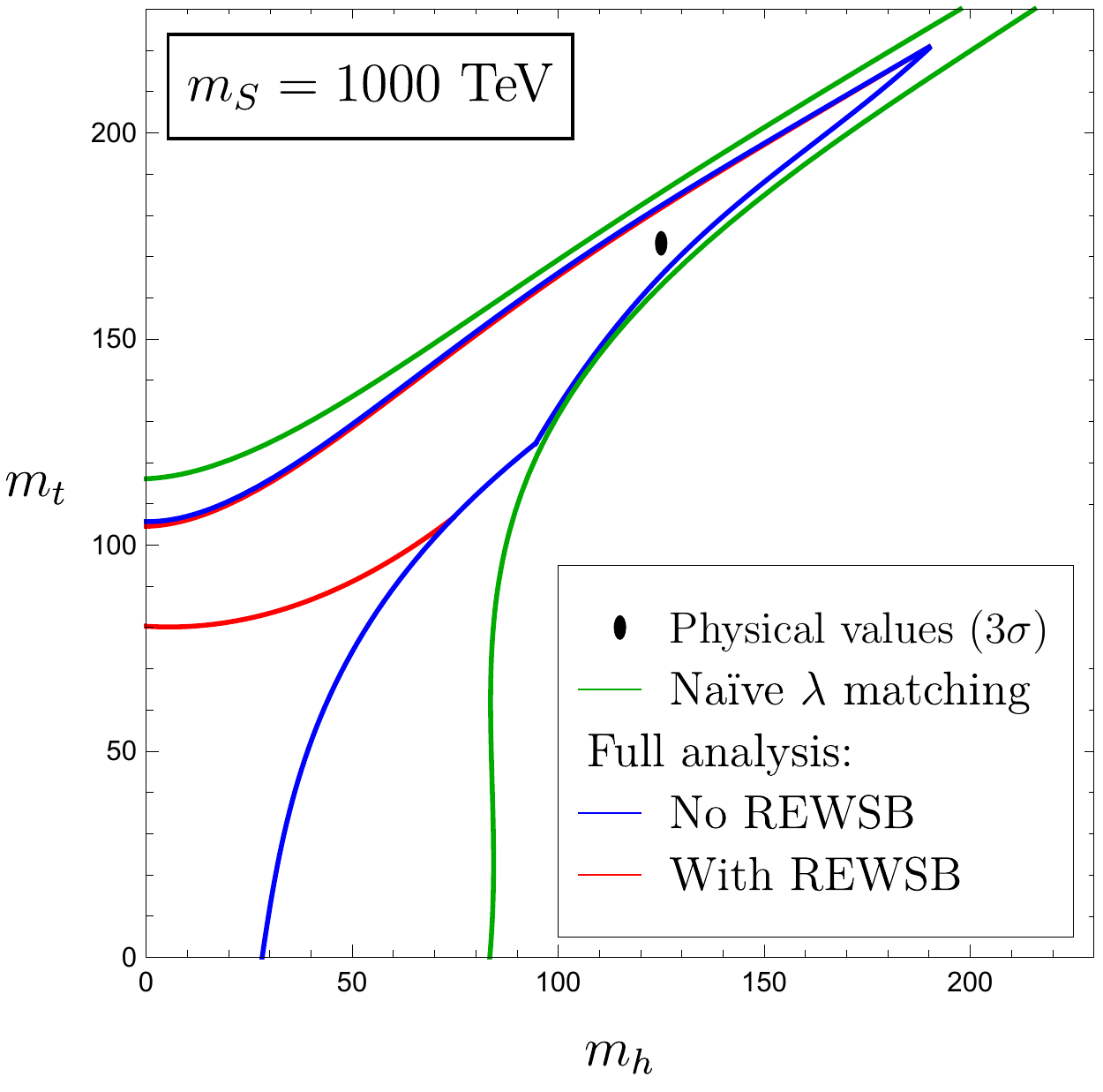}
			\end{minipage}
			}
			\caption{Na\"ive viable region (green curves), viable region with (red curves) and without (blue
								 curves) REWSB, for $\mS=100 \TeV$ (left plot) and $\mS=1000 \TeV$ (right plot). 
			\label{figF_100&1000TeV_BRG}}
\end{figure}


The top-left plot of Figure \ref{figF_5TeV_comparison} shows how the viable region expands when the boundary conditions at $\mG$ are scanned more widely (precisely, in \eq(\ref{eqF_b_gutBCs}) we switch from $\bar r=3$ to $\bar r=10$). As can be seen, the upper and right borders expands (as expected), but only very mildly.
On the other hand, the variation is quite pronounced for the lower border. The reason for this effect has already been discussed. More explicitly, for 
low  $\mt$ values electroweak symmetry breaking can be achieved allowing a bigger (but somehow unnatural) splitting between $\mHu^2(\mG)$ and $\mHd^2(\mG)$.

The impact of removing configurations leading to a Landau pole for $\yt(\mu)$ before the GUT scale, and the impact of tigther constraints on the higher-order 
terms in the $\lambda$-matching condition, are illustrated in the top-right and bottom-left plots of Figure \ref{figF_5TeV_comparison}, respectively. 
Finally, in the bottom-right plot of Figure \ref{figF_5TeV_comparison}  we show the shift in the shape of the 
 viable region  due to a reduction of the maximal $\tan\bS$ value considered.

\subsubsection{The Radiative EWSB condition at low $\mS$}

Still considering the low  $\mS$ scenario (with $\mS=5$ TeV), we can now examine how the viable region shrinks when the additional requirement of REWSB, 
i.e.~the simultaneous requirement of \eq(\ref{eqE_b_vacReq2}) and  \eq(\ref{eqF_b_REWSB}),
is imposed. Naïvely, one should expect the REWSB condition to be more difficult to be achieved in the region where 
the RG effects induced by $\yt(\mu)$ are smaller, hence in the low-$\mt$ region. This is indeed what happens, as shown in Figure~\ref{figF_5TeV_BRG}.

As can be seen, the main effect of the REWSB condition is to reduce significantly the lower border of the allowed region, cutting out the 
low-$\mt$ region: for $\mt\approx 80 ~\GeV$ the red border (denoting the constraint obtained imposing REWSB) 
detaches itself from the blue one (obtained without this extra conidition). 
The net effect is not a sharp (horizontal) cut in $\mt$, since the decrease in $\yt(\mu)$  can be compensated by
an appropriate change in $\mHud^2$. However, the latter makes the $\lambda$-matching harder to achieve, resulting into the non-trivial 
slope in $\mt$ vs.~$\mh$ exhibited by the red curve in Figure \ref{figF_5TeV_BRG}. 
Extrapolating the curve to $\mh \to 0$, we deduce the  
existence of a critical value for $\mt$ ($\mt^{\rm min} \sim 50 ~\GeV$ for $\mS = 5 ~\TeV$) below which REWSB cannot be achieved
independently of the other parameters.

\subsubsection{High-scale SUSY}

We now  proceed analysing how the allowed region change by raising the soft-breaking scale. 
In Figure \ref{figF_100&1000TeV_BRG} we show the result obtained for $\mS=100~\TeV$ (left plot) and $\mS=1000~\TeV$ (right plot),
which should be compared with the $\mS=5~\TeV$ case shown in Figure~\ref{figF_5TeV_BRG}.

On general grounds, the main effect of raising $\mS$ is a longer evolution of the Higgs self coupling following SM RG equations.
In the large $\mt$ region, where $\lambda_\sSM(\mu)$ tends to become negative, this imply a significant 
shrinking of the allowed region in order to avoid too negative $\lambda_\sSM(\mS)$ values. More precisely, 
we expect a significant lowering of the upper borders and a less pronounced lowering of the lower borders.
At the same time, raising $\mS$ we move forward the matching scale for $\yt$. This implies 
a reduced the influence of $\yt(\mu)$ in the MSSM RG flow, in turn, shrink the region where RWESB can occur. 
Last but not least, by raising $\mS$ the impact of threshold corrections in the $\lambda$-matching condition is significantly reduced.

\FPeval{\nA}{0.42}		\FPeval{\nB}{0.98}
\begin{figure}[t]
			\centering
			\begin{minipage}{\nA\textwidth}
						\centering
						\includegraphics[width=\nB\textwidth]{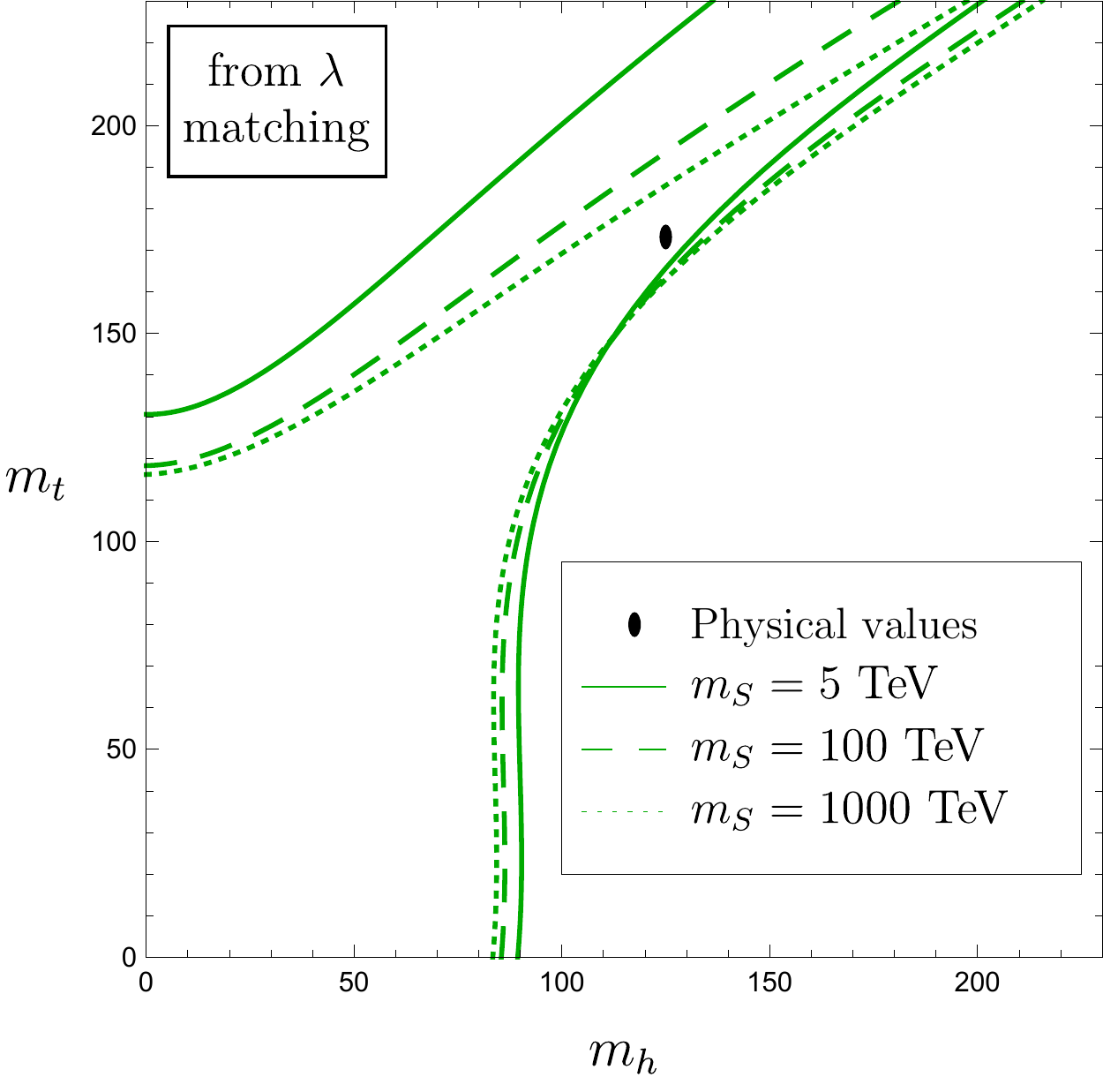}
			\end{minipage}
			~~~~~
			\begin{minipage}{\nA\textwidth}
						\centering
						\includegraphics[width=\nB\textwidth]{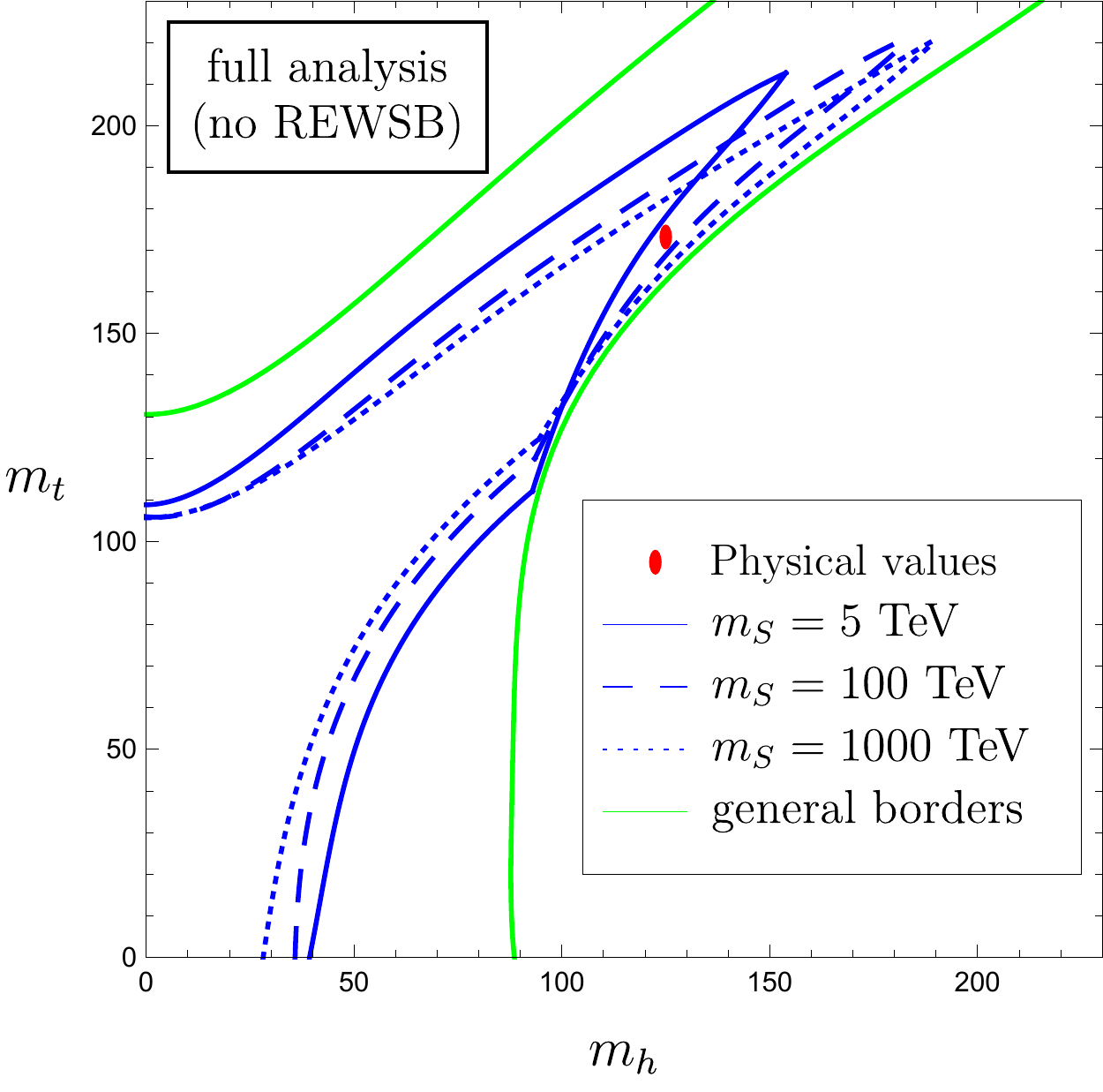}
			\end{minipage}
			\\ ~ \phantom{a} ~ \\
			\begin{minipage}[b]{\nA\textwidth}
						\centering
						\includegraphics[width=\nB\textwidth]{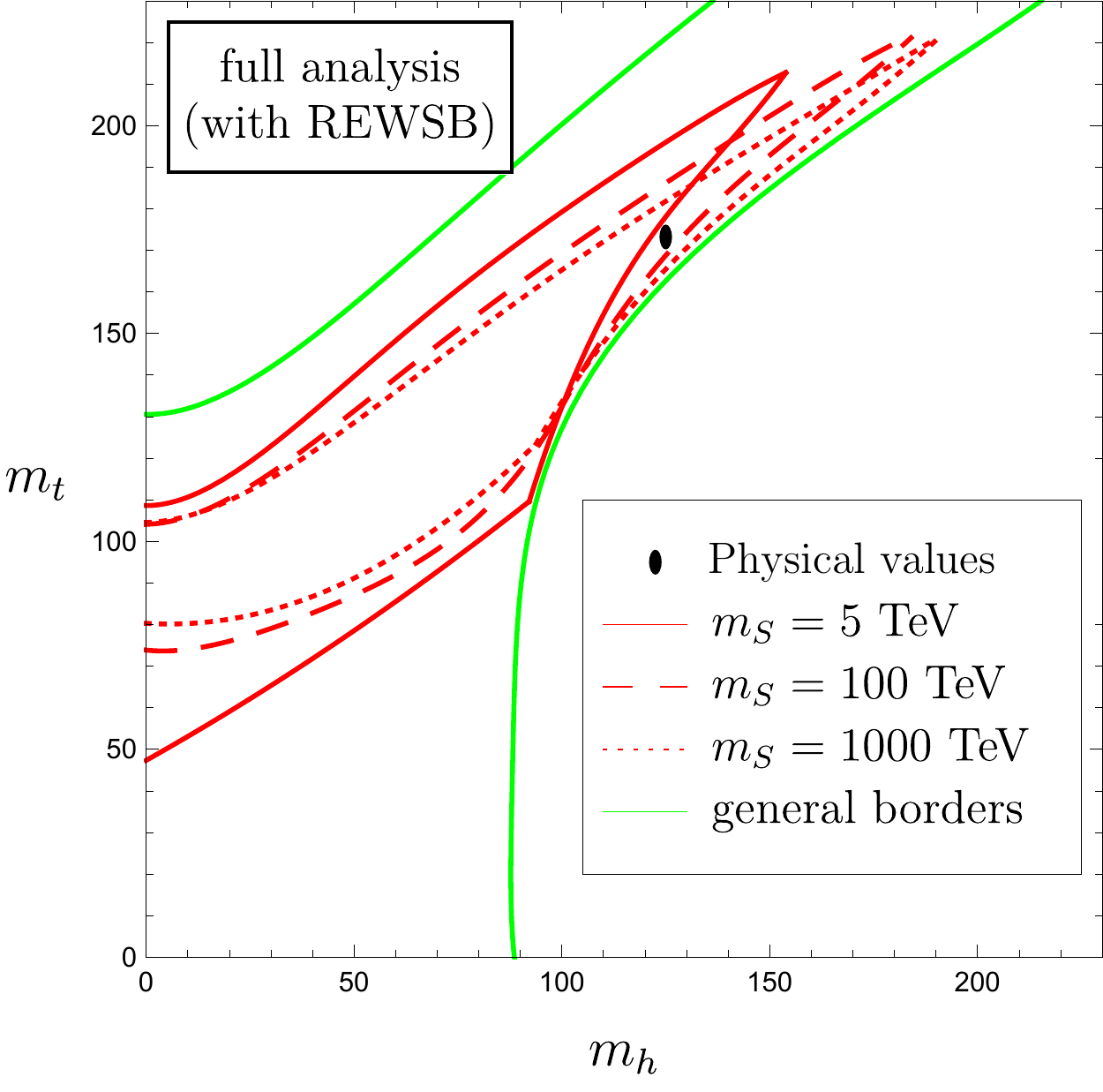}
			\end{minipage}
			~~~~~
			\begin{minipage}[b]{\nA\textwidth}
			\caption{Comparison of viable regions for three different $\mS$ values: 
								$\mS=5\TeV$ (solid line), $\mS=100\TeV$ (dashed line), and $\mS=1000\TeV$ (dotted line). 
								In all plots, the black dot denotes the physical  values of $\mh$ and $\mt$.
								\emph{Top-left plot}: comparison of the na\"ive viable regions.
								\emph{Top-right (bottom-left) plot}: comparison of the viable regions without (with) the 
								optional REWSB condition. 
			\label{figF_mS_comparison}} 
			\end{minipage}
\end{figure}

In Figure \ref{figF_mS_comparison} we present a detailed comparison of the  various independent 
constraints for $\mS=\{5, 100, 1000\}$ TeV.   
 As can be easily understood, the change is particularly pronounced 
for the green curves corresponding to the na\"ive $\lambda$-matching, that only depend on the value 
of $\lambda_\sSM(\mS)$.  The second largest effect is the reduction of the low $\mt$ region after 
imposing the REWSB condition. 
A point worth to note is that the physical point is always well within the allowed region, 
even if the size of the allowed region shrinks significantly for larger $\mS$ values. 

\subsection{A comparison with the SM metastability region}

\begin{figure}
\center
\includegraphics[width=0.5\textwidth]{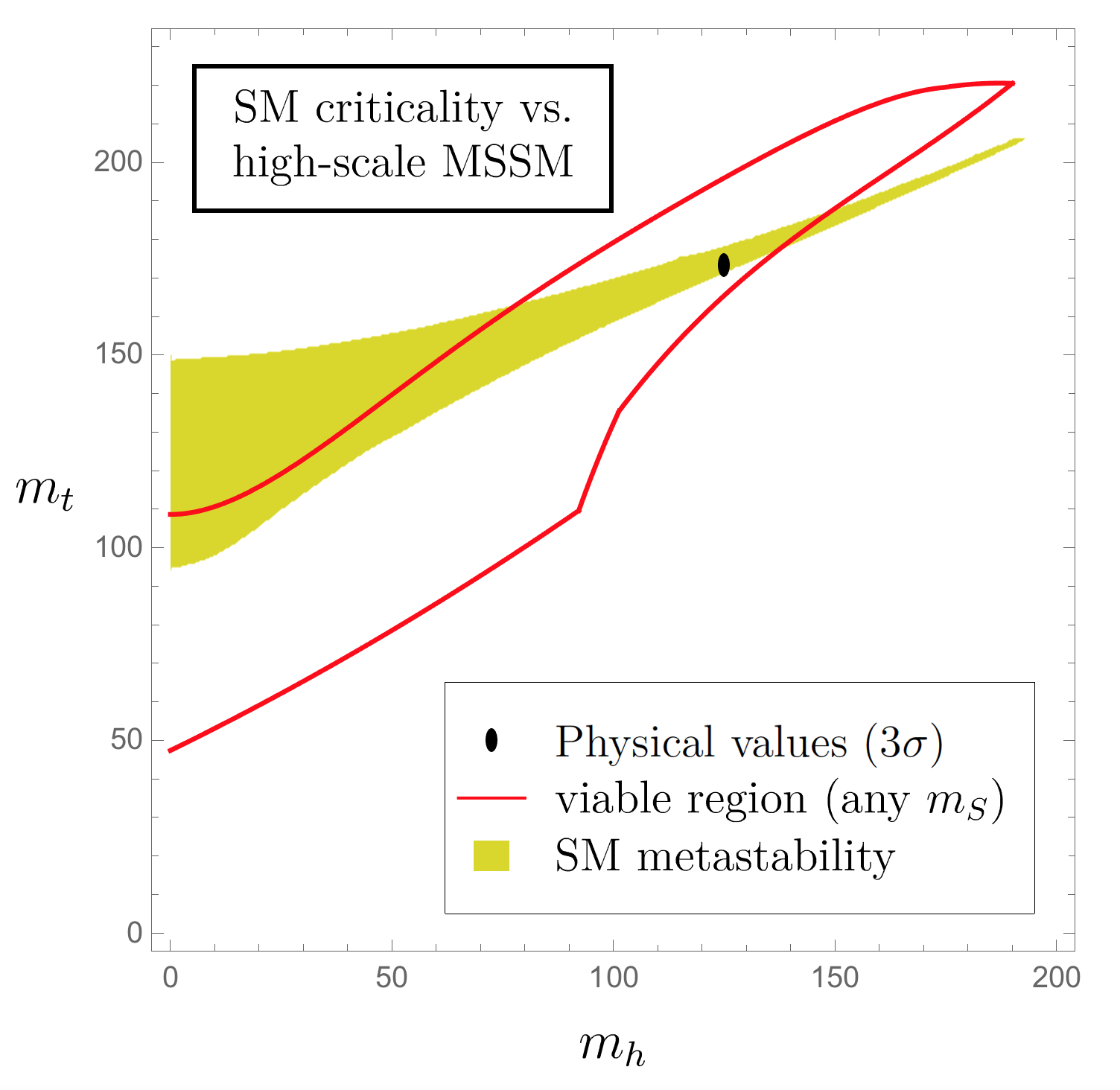}
\caption{Comparison, in the $(\mh,\mt)$ plane, between the MSSM viable region (for any $\mS$)
and the Standard Model   metastability region. \label{figF_SMvsMSSM}}
\end{figure}

As final step of our discussion, we compare the MSSM predictions in the $(\mh,\mt)$  plane,
scanning the full allowed range of free parameters, with the predictions for the same set of observables obtained extrapolating the validity of the SM up to the Planck scale. 

The prediction of  $\mh$ vs.~$\mt$, obtained extrapolating the validity of the SM up to very high energy scales,
has been extensively discussed in the literature (see e.g.~\cite{Casas:1994qy,Altarelli:1994rb,Isidori:2001bm,ArkaniHamed:2008ym,Bezrukov:2009db,Ellis:2009tp,Isidori:2007vm,Degrassi:2012ry,Buttazzo:2013uya,Bezrukov:2012sa} 
and references therein). As shown by the recent precise analyses performed after the Higgs mass 
discovery~\cite{Degrassi:2012ry,Buttazzo:2013uya,Bezrukov:2012sa},  the experimental values of $\mh$ and $\mt$ exhibit a remarkable feature:
they lie within, and quite close to the border, of the narrow SM metastability region. The latter is the region where the SM potential, extrapolated up to high field values,  
is not stable (under the hypothesis of no new physics below the Planck mass) but is sufficiently metastable 
(assuming no further destabilizations induced at or around the Planck scale). 
This near-critical configuration  might have very interesting cosmological implications (see e.g.~\cite{ArkaniHamed:2008ym,Isidori:2007vm,Bezrukov:2014bra,Espinosa:2015qea}), and it has been advocated as an argument in favour of the ``multiverse'', where such 
near-critical configuration might emerge as an attractor~\cite{Buttazzo:2013uya}.

Through the study carried out in this paper, we are now in the position to make a claim of similar nature for the MSSM 
(and somehow more loosely for supersymmetric theories in general).
Combining the viable regions for different $\mS$ values (and requiring REWSB), we can infer an overall MSSM viable region.
This region is shown in Figure \ref{figF_SMvsMSSM}, where it is compared to the SM metastability region.

The supersymmetric viable region in Figure \ref{figF_SMvsMSSM} encloses all points which admit a MSSM parameter configuration satisfying 
the four conditions illustrated in Section~\ref{secF_method}, namely gauge coupling unification, the 
$\lambda$-matching condition, a natural range for the soft breaking  terms at the GUT scale, 
and radiative EWSB.   As it can be seen, somehow quite surprisingly, 
such region is not much bigger than the SM metastability region, and the 
experimental values of  $\mh$ and $\mt$  lies well within it. 

It is worth stressing that a key role in restricting the 
size of the SUSY-allowed region is played by the Radiative EWSB condition (a condition that is often ignored
when predicting the Higgs mass in SUSY frameworks), which is essential in cutting out low $m_t$ values. 
It is also worth noting that Figure~\ref{figF_SMvsMSSM} only assumes a lower bound on $\mS$ of a few TeV 
(the upper bound follows from the request of gauge coupling unification). 
If one could fix the $\mS$ value in the high-scale region ($100~{\rm TeV} < \mS < 1000~{\rm TeV}$), 
the resulting   SUSY-allowed region would be further restricted (see Fig.~\ref{figF_100&1000TeV_BRG}) with the 
experimental values of  $\mh$ and $\mt$ still well within it.

\section{Conclusions}
Given the theoretical appeal of the MSSM  as possible ultraviolet completion of the Standard Model, 
in this paper we have tried to address the question of how likely is to reproduce the 
observed SM spectrum, and the absence of new-physics signals, starting from a generic MSSM.  
On the absence of new-physics signals we have little to say: we simply assumed 
this is a consequence of a minimum value for the soft breaking scale exceeding a few TeV. 
Given this first assumption, and the second key assumption that gauge couplings and 
Fermi scale have to be the observed ones (an hypothesis that can be justified using 
anthropic considerations), we  analysed  which are the 
natural predictions for top and Higgs masses in a generic MSSM.
In order to restrict the parameter space of the underlying model, we have further assumed 
gauge-coupling unification, radiative electroweak symmetry breaking, and a natural (non-splitted) spectrum 
of soft-breaking terms at the GUT scale. Three hypotheses that add theoretical appeal to the MSSM 
as ultraviolet completion of the Standard Model. 

The result of our analysis is summarised by the plot in Fig.~\ref{figF_SMvsMSSM}, where we 
compare the SUSY-viable region in the $(\mh,\mt)$ plane with the SM metastability region, 
obtained extrapolating the validity of the SM up to very high energy scales. 
As it can be seen, the SUSY-allowed region is not much bigger than the SM metastability region, and the 
experimental values of  $\mh$ and $\mt$  lies well within it. While it is certainly too strong to state that this is an ``evidence'' in flavour of 
a supersymmetric extension of the MSSM, we argue that such result is not less surprising that the near-criticality observed in the SM case.
As such, this observation is suitable for speculations similar to those proposed in the SM case, such as the emergence of the observed 
SM spectrum from a supersymmetric multiverse (with soft-breaking scale ranging from about 1 TeV up to the maximal value compatible with 
gauge-coupling unification).

\subsection*{Acknowledgements}

This research was supported in part by the Swiss National Science Foundation (SNF) under contract 200021-159720.

\renewcommand\thesection{\Alph{section}}
\setcounter{section}{0}

\section{Numerical value of $\ltr$ }
\label{sect:appA}

\begin{figure}[t]
			\centering
			\mbox{
			\begin{minipage}[b]{0.45\textwidth}
						\centering
						\includegraphics[width=0.98\textwidth]{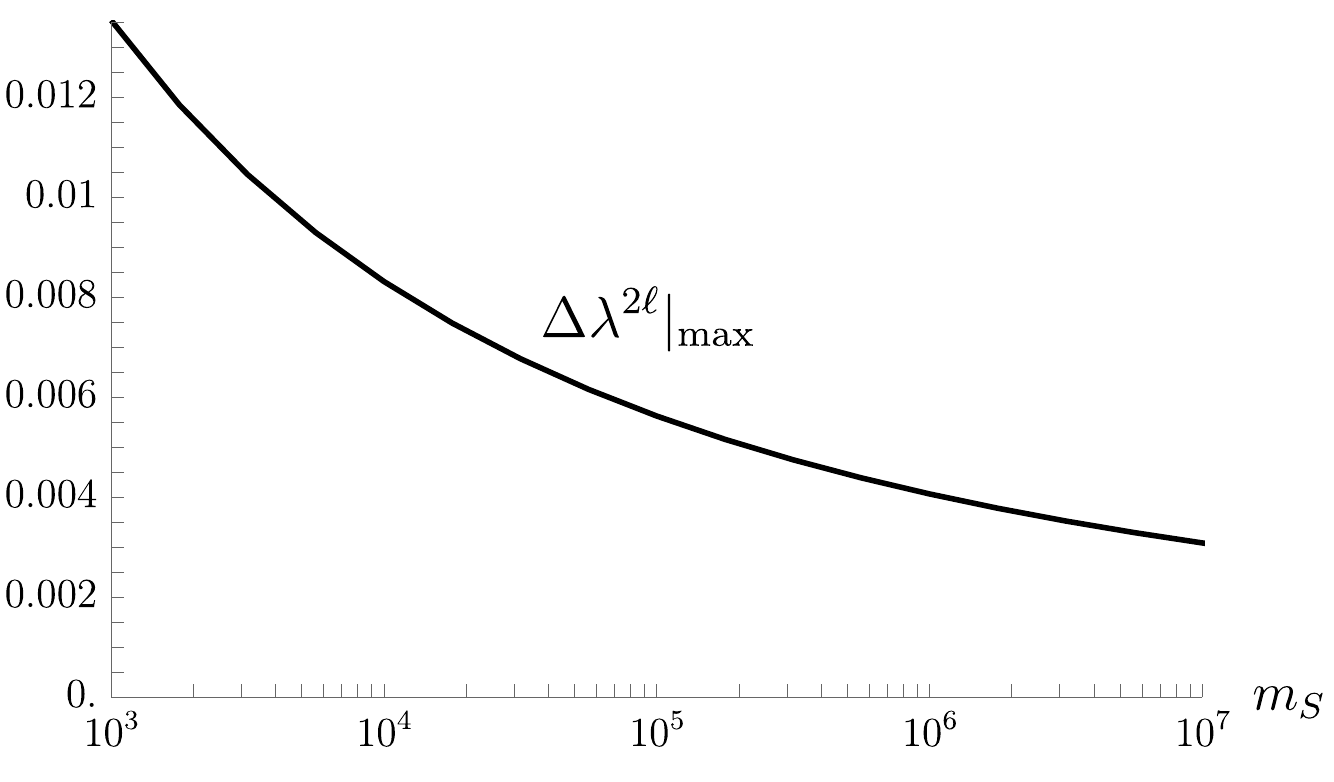}
						\caption{Maximal value of $\Delta\lambda^{2\ell oop}$ as a function of $\mS$. 
						\label{figF_lam2loop}}
			\end{minipage}
			~~~
			\begin{minipage}[b]{0.4\textwidth}
						\centering
						\begin{tabular}{| c | c |}
									\hline	$\mS$					&		$\ltr$ 			\\
									\hline	$5$ TeV				&		$0.01$			\\
									\hline	$100$ TeV			&		$0.006$		\\
									\hline	$1000$ TeV 		&		$0.004$		\\
									\hline	\multicolumn{2}{c}{~~}
						\end{tabular}
						\captionof{table}{Selected values of $\ltr$ for different $\mS$ values used in our analysis.
						\label{tabF_lamThr}}
			\end{minipage}
			}
\end{figure}

Here we briefly summarise the procedure adopted to define the numerical error in the $\lambda$-matching condition ($\ltr$).
Since $\lambda_\sSM$ renormalise in a non-multiplicative fashion, we have fixed the maximal value of the difference 
$|\lambda_\sSM(\mS)-\lambda_\MSSM(\mS)|$, rather than a possible error on the ratio 
$|\lambda_\sSM(\mS)/\lambda_\MSSM(\mS)|$.
On the other hand, given this difference has a sizeable dependence on the overall scale of the soft spectrum, 
we have varied it as a function of  $\mS$.
As an estimator of such difference we have used the leading two-loop contributions to $\lambda_\MSSM(\mS)$ reported in
\Ref\cite{Bagnaschi:2014rsa}.
The maximal values of $\Delta\lambda^{2\ell oop}$ for different supersymmetry scales are plotted in Figure \ref{figF_lam2loop}.
The numerical values of $\ltr$   extracted by this procedure for $\mS=\{ 5, 100, 1000\}$~TeV
(the reference $\mS$ values adopted in the analysis) 
are summarised in Table \ref{tabF_lamThr}.

\begin{thebibliography}{99}
{\footnotesize


\bibitem{Ellis:1990wk}
  J.~R.~Ellis, S.~Kelley and D.~V.~Nanopoulos,
  Phys.\ Lett.\ B {\bf 260} (1991) 131.


\bibitem{Amaldi:1991cn}
  U.~Amaldi, W.~de Boer and H.~Furstenau,
  Phys.\ Lett.\ B {\bf 260} (1991) 447.


\bibitem{Langacker:1991an}
  P.~Langacker and M.~x.~Luo,
  Phys.\ Rev.\ D {\bf 44} (1991) 817.



\bibitem{Ellis:1990nz}
  J.~R.~Ellis, G.~Ridolfi and F.~Zwirner,
  Phys.\ Lett.\ B {\bf 257} (1991) 83.


\bibitem{Haber:1993an}
  H.~E.~Haber and R.~Hempfling,
  Phys.\ Rev.\ D {\bf 48} (1993) 4280
  [hep-ph/9307201].
  

\bibitem{Carena:1995wu}
  M.~Carena, M.~Quiros and C.~E.~M.~Wagner,
  Nucl.\ Phys.\ B {\bf 461} (1996) 407
  [hep-ph/9508343].


\bibitem{Heinemeyer:1998yj}
  S.~Heinemeyer, W.~Hollik and G.~Weiglein,
  Comput.\ Phys.\ Commun.\  {\bf 124} (2000) 76
  [hep-ph/9812320].


\bibitem{ArkaniHamed:2004fb}
  N.~Arkani-Hamed and S.~Dimopoulos,
  JHEP {\bf 0506} (2005) 073
  [hep-th/0405159].


\bibitem{Giudice:2004tc}
  G.~F.~Giudice and A.~Romanino,
  Nucl.\ Phys.\ B {\bf 699} (2004) 65
  [hep-ph/0406088].


\bibitem{Hall:2009nd}
  L.~J.~Hall and Y.~Nomura,
  JHEP {\bf 1003} (2010) 076
  [arXiv:0910.2235].

\bibitem{Giudice:2011cg}
  G.~F.~Giudice and A.~Strumia,
  Nucl.\ Phys.\ B {\bf 858} (2012) 63
  [arXiv:1108.6077].


\bibitem{Arvanitaki:2012ps}
  A.~Arvanitaki, N.~Craig, S.~Dimopoulos and G.~Villadoro,
  JHEP {\bf 1302} (2013) 126
  [arXiv:1210.0555].


\bibitem{Bagnaschi:2014rsa}
  E.~Bagnaschi, G.~F.~Giudice, P.~Slavich and A.~Strumia,
  JHEP {\bf 1409} (2014) 092
  [arXiv:1407.4081].


\bibitem{Vega:2015fna}
  J.~Pardo Vega and G.~Villadoro,
  JHEP {\bf 1507} (2015) 159
  [arXiv:1504.05200].

\bibitem{Ellis:2017erg}
  S.~A.~R.~Ellis and J.~D.~Wells,
  Phys.\ Rev.\ D {\bf 96} (2017) no.5,  055024
  [arXiv:1706.00013].





\bibitem{Degrassi:2012ry}
  G.~Degrassi, S.~Di Vita, J.~Elias-Miro, J.~R.~Espinosa, G.~F.~Giudice, G.~Isidori and A.~Strumia,
  JHEP {\bf 1208} (2012) 098
  [arXiv:1205.6497].

\bibitem{Bezrukov:2012sa}
  F.~Bezrukov, M.~Y.~Kalmykov, B.~A.~Kniehl and M.~Shaposhnikov,
  JHEP {\bf 1210} (2012) 140
  [arXiv:1205.2893].

\bibitem{Buttazzo:2013uya}
  D.~Buttazzo, G.~Degrassi, P.~P.~Giardino, G.~F.~Giudice, F.~Sala, A.~Salvio and A.~Strumia,
  JHEP {\bf 1312} (2013) 089
  [arXiv:1307.3536].


\bibitem{Casas:1994qy}
  J.~A.~Casas, J.~R.~Espinosa and M.~Quiros,
  Phys.\ Lett.\ B {\bf 342} (1995) 171
  [hep-ph/9409458].

\bibitem{Altarelli:1994rb}
  G.~Altarelli and G.~Isidori,
  Phys.\ Lett.\ B {\bf 337} (1994) 141.

\bibitem{Isidori:2001bm}
  G.~Isidori, G.~Ridolfi and A.~Strumia,
  Nucl.\ Phys.\ B {\bf 609} (2001) 387
  [hep-ph/0104016].
 

\bibitem{Bezrukov:2009db}
  F.~Bezrukov and M.~Shaposhnikov,
  JHEP {\bf 0907} (2009) 089 
  [arXiv:0904.1537].
 
\bibitem{Ellis:2009tp}
  J.~Ellis, J.~R.~Espinosa, G.~F.~Giudice, A.~Hoecker and A.~Riotto,
  Phys.\ Lett.\ B {\bf 679} (2009) 369 
  [arXiv:0906.0954].
 
 
\bibitem{Isidori:2007vm}
  G.~Isidori, V.~S.~Rychkov, A.~Strumia and N.~Tetradis,
  Phys.\ Rev.\ D {\bf 77} (2008) 025034 
  [arXiv:0712.0242].
 
\bibitem{ArkaniHamed:2008ym}
  N.~Arkani-Hamed, S.~Dubovsky, L.~Senatore and G.~Villadoro,
  JHEP {\bf 0803} (2008) 075
  [arXiv:0801.2399].


\bibitem{Bezrukov:2014bra}
  F.~Bezrukov and M.~Shaposhnikov,
  Phys.\ Lett.\ B {\bf 734} (2014) 249
  [arXiv:1403.6078]. 

\bibitem{Espinosa:2015qea}
  J.~R.~Espinosa, G.~F.~Giudice, E.~Morgante, A.~Riotto, L.~Senatore, A.~Strumia and N.~Tetradis,
  JHEP {\bf 1509} (2015) 174
  [arXiv:1505.04825].


}


\end{thebibliography}
\end{document}